\documentclass[12pt]{iopart}

\usepackage[dvips]{graphicx}
\usepackage{mathptmx}

\begin{document}

\title[Moment hierarchy from quantum kinetic theory]{Fluid moment hierarchy equations derived from gauge invariant quantum kinetic theory}

\author{F. Haas, J. Zamanian, M. Marklund, and G. Brodin}

\address{Department of Physics, Ume{\aa} University, SE\,--\,901 87 Ume{\aa}, Sweden}

\begin{abstract}
The gauge invariant electromagnetic Wigner equation is taken as the basis for a fluid-like system describing quantum plasmas, derived from the moments of the gauge invariant Wigner function. The use of the standard, gauge dependent Wigner function is shown to produce inconsistencies, if a direct correspondence principle is applied. The propagation of linear transverse waves is considered and shown to be in agreement with the kinetic theory in the long wavelength approximation, provided an adequate closure is chosen for the macroscopic equations. A general recipe to solve the closure problem is suggested.
\end{abstract}
%\PACS{05.60.Gg, 52.35.Hr, 67.10.Jn}
%\maketitle

\section{Introduction}
The Wigner function is the quantum equivalent of the classical particle distribution function and can be used to calculate average values of physical observables \cite{Wigner}. In most cases, the time evolution of the Wigner function is evaluated considering only scalar potentials, hence 
without the inclusion of magnetic fields. One reason for this is the considerable analytic complexity of the electromagnetic Wigner equation. Indeed even the electrostatic Wigner equation already is a cumbersome integro-differential equation which hardly can be examined except in the linear limit. However, the emergence of new areas like spintronics \cite{Zutic} where magnetic effects are crucial makes it desirable to have quantum kinetic models allowing for nonzero vector potentials. In this situation the gauge invariance of the Wigner function should be assured from the very beginning in order to avoid inconsistencies, a point somewhat neglected in previous studies. It is the purpose of this work to stress the relevance and properties of the gauge invariant Wigner function (GIWF) \cite{Stratonovich, Serimaa,Levanda} in connection with quantum plasmas problems. In addition we provide a macroscopic (moments) formulation starting from the electromagnetic Wigner-Maxwell system, substantially generalizing the recently introduced moments system derived from the Wigner-Poisson equations \cite{Haas}. The resulting macroscopic equations are a step toward the inclusion of spin dependent variables, postponed to future considerations.  

The advantages in macroscopic formulations are in their relative simplicity, so that the nonlinear regimes are not necessarily unaccessible apart from numerical simulations. Notice however that our fluid approach does not imply any fluid approximations, in the sense that we are not supposing a large collision rate or a short mean free path for instance. If we are interested only in basic quantities like particle, current or energy densities, nothing forbids to compute moments of the Wigner function in order to derive fluid-like equations for the time-evolution of these variables. The roots of the moments descriptions in plasma theory can be traced back to Grad \cite{Grad}. The price of replacing the more detailed kinetic models by macroscopic models is the loss of information on kinetic phenomena like Landau damping, the plasma echo and many others.

This work is organized as follows. In Section II we briefly review the definition and properties of the GIWF. Section III develop the corresponding fluid moment hierarchy equations. Section IV consider the propagation of transverse waves and the closure problem in this case. Section V is dedicated to the conclusions. In addition, it is included the Appendix A where the closure of the fluid-like system is discussed.

\section{Basic properties of the gauge invariant Wig\-ner func\-tion}
A sensible definition of gauge invariant one-particle Wigner function $f = f({\bf r},{\bf v},t)$ was introduced by Stratonovich \cite{Stratonovich}, and rediscovered by Irving \cite{Irving}. Since in this work we are not concerned with relativistic phenomena, we write it in a non-covariant form,  
\begin{eqnarray}
&&
f({\bf r},{\bf v},t) = \left(\frac{m}{2\pi\hbar}\right)^{3}\,\int d{\bf s}\,\exp\left[\frac{i{\bf s}}{\hbar}\cdot\left(m{\bf v} + q\int_{-1/2}^{1/2}d\tau{\bf A}({\bf r} + \tau{\bf s},t)\right)\right] 
\nonumber \\ \label{e1} 
&& \qquad\qquad\quad \times \psi^{*}\left({\bf r}+\frac{\bf s}{2},t\right)\psi\left({\bf r} - \frac{\bf s}{2},t\right) \,,
\end{eqnarray}
where ${\bf r}$ and ${\bf v}$ are the position and velocity vectors and $t$ the time. The wave function is assumed to be normalized to unity. In addition, $\hbar$ is Planck's constant divided by $2\pi$, ${\bf A}({\bf r}, t)$ is the vector potential, $m$ and $q$ are the mass and charge of a particle in a pure state described by a wave function $\psi({\bf r},t)$. The properties to be discussed in this Section hold equally well in the case of mixed states. In contrast to the original definition of Wigner function \cite{Wigner} via the canonical momentum, the object $f$ in Eq. (\ref{e1}) is written in terms of the kinetic momentum $m{\bf v}$. The extra integral in Eq. (\ref{e1}) containing the vector potential compensates for the change in the wave function in a local gauge transformation. The use of a non-covariant, one-time pseudo-distribution renders the interpretation issues of $f$ less obscure than in a four-dimensional space-time version, as stressed in Ref. \cite{Bia}.

Naturally there are other ways to obtain GIWFs, {\it e.g.} through certain path integrals involving the vector potential \cite{Carruthers}. However, the phase factor in Eq.\ (\ref{e1}) can be justified \cite{Levanda} in terms of the minimal coupling principle. Moreover, as discussed in more detail elsewhere, the function of the phase factor is to convert any gauge into the axial gauge \cite{Levanda}. For our purposes, the choice of the form (\ref{e1}) is due to convenience as it provides a non-ambiguous way to calculate averaged quantities. If instead one take a GIWF in terms of a line integral $\int_{{\bf r}_1}^{{\bf r}_2}\,{\bf A}({\bf s},t)\cdot d{\bf s}$ one introduce the further difficulty of the choice of integration path from ${\bf r}_1$ to ${\bf r}_2$ (cf. Eq. (2.157) of Ref. \cite{Carruthers}). 

The properties of the GIWF have been detailed in \cite{Serimaa, Levanda}. Nevertheless for completeness we discuss some of them once again. From $f$ we can compute the very basic zeroth and first order moments
\begin{eqnarray}
\label{e2}
\int\,d{\bf v}\,f &=& |\psi|^2 \,,\\
\label{e3}
\int\,d{\bf v}\,{\bf v}\,f &=& \frac{i\hbar}{2\,m}(\psi\nabla\psi^{*} - \psi^{*}\nabla\psi) - \frac{q}{m}|\psi|^2\,{\bf A} \,,
\end{eqnarray}
with the interpretation of particle and current densities respectively. By construction these quantities are invariant under the local gauge transformation
\begin{equation}
\label{e4}
{\bf A} \rightarrow {\bf A} + \nabla\Lambda \,, \quad \psi \rightarrow \psi\exp\left(\frac{i\,q\,\Lambda}{\hbar}\right) \,,
\end{equation}
where $\Lambda = \Lambda({\bf r},t)$ is an arbitrary differentiable function. 

If the starting point is the usual (gauge dependent) Wigner function 
\begin{equation}
\label{e5}
f^{\,GD}({\bf r},{\bf p},t) = \frac{1}{(2\pi\hbar)^3}\int d{\bf s}\,\exp\left(\frac{i\,{\bf p}\cdot{\bf s}}{\hbar}\right) \psi^{*}\left({\bf r}+\frac{\bf s}{2},\,t\right)\psi\left({\bf r} - \frac{\bf s}{2},\,t\right) \,,
\end{equation}
which is written in terms of the canonical momentum ${\bf p} = m\,{\bf v} - q\,{\bf A}$, one obtains gauge independent results for the zeroth, first and second order moments, but gauge dependent quantities when considering higher order moments. Of course implicitly we assume that all physical objects should be gauge independent. Serious discrepancies occurs when calculating the evolution equation for the second order moment of the usual Wigner function and the GIWF, as will be shown in the next Section. In all cases it is safer to work with $f$ as given in Eq. (\ref{e1}).

The time-evolution of the GIWF was considered already by Stratonovich \cite{Stratonovich}, but a particularly illuminating form to express it was provided by Serimaa {\it et al.} \cite{Serimaa} according to 
\begin{equation}
\label{e11}
\left\{\frac{\partial}{\partial t} + ({\bf v} + \Delta\tilde{\bf v})\cdot\frac{\partial}{\partial{\bf r}} + \frac{q}{m}\,\left[\tilde{\bf E} + ({\bf v} + \Delta\,\tilde{\bf v})\times\tilde{\bf B} \right]\,\cdot\frac{\partial}{\partial{\bf v}}
\right\}\,f({\bf r},{\bf v},t) = 0 \,.
\end{equation}
Here, we introduced the operators 
\begin{eqnarray}
\label{e8}
\Delta\tilde{\bf v} &=& \frac{i\,\hbar\,q}{m^2}\,\frac{\partial}{\partial{\bf v}}\times\int_{-1/2}^{1/2}d\tau\,\tau{\bf B}\left({\bf r}+\frac{i\,\hbar\,\tau}{m}\,\frac{\partial}{\partial{\bf v}},\,t\right) \,,\\ 
\label{e9}
\tilde{\bf E} &=& \int_{-1/2}^{1/2}d\tau\,{\bf E}\left({\bf r}+\frac{i\,\hbar\,\tau}{m}\,\frac{\partial}{\partial{\bf v}},\,t\right) \,,\\
\label{e10}
\tilde{\bf B} &=& \int_{-1/2}^{1/2}d\tau\,{\bf B}\left({\bf r}+\frac{i\,\hbar\,\tau}{m}\,\frac{\partial}{\partial{\bf v}},\,t\right) \,,
\end{eqnarray}
where ${\bf B} = {\bf B}({\bf r}, t)$ and ${\bf E} = {\bf E}({\bf r}, t)$ are the magnetic and electric fields respectively. The kinetic equation (\ref{e11}) follows from the Schr\"odinger equation for the wave function or, alternatively, from the von Neumann equation solved by the density matrix. 

As apparent from Eq.\ (\ref{e11}), the kinetic equation satisfied by $f$ is formulated in terms of the physical fields, unlike the equation solved by $f^{\,GD}$ which is written in terms of the scalar and vector potentials \cite{Haas2} and which can be shown to be {\it not} gauge invariant, a serious drawback. Moreover Eq.\ (\ref{e11}) is almost in the form of a Vlasov equation, with two differences: the electromagnetic fields are replaced by $\tilde{\bf E}$ and $\tilde{\bf B}$ defined in Eqs.\ (\ref{e9})--(\ref{e10}); \, the velocity vector is displaced by the intrinsically quantum mechanical perturbation $\Delta\tilde{\bf v}$ defined in Eq.\ (\ref{e8}). Notice that this perturbation $\Delta\tilde{\bf v}$ vanishes in the electrostatic case. In calculating Eqs.\ (\ref{e8})--(\ref{e10}), it is assumed that the electromagnetic fields are analytic, so that the integrals are evaluated after Taylor expanding and then replacing ${\bf r}$ by the indicated argument ${\bf r} + i\hbar\,(\tau/m)\,\partial/\partial{\bf v}$. A further difference in comparison to the Vlasov equation is that not any function $f$ on phase space can be taken as a Wigner function. Too spiky functions violating the uncertainty principle should be ruled out. And, of course, the Wigner function is not strictly a probability distribution, since in general it is negative in certain regions of phase space. 

To sum up the pseudo-distribution in Eq.\ (\ref{e1}) provides a practical and non-ambiguous recipe for a GIWF and Eq. (\ref{e11}) is the associated kinetic equation. In the next Section we derive a system of partial differential equations satisfied by macroscopic quantities obtained taking moments of the GIWF.  

\section{Fluid moments hierarchy}
In spite of the apparent simplicity, actually Eq.\ (\ref{e11}) becomes quite complicated after developing the operators $\tilde{\bf E}$ and $\tilde{\bf B}$. In practice, nonlinear problems are unaccessible in this formulation, specially remembering that the electromagnetic field should be self-consistently determined through Maxwell equations. Hence apart from linear problems this Wigner-Maxwell system can be helpful only by means of numerical simulations, which are themselves not evident due to the complexity of the system. This motivate the creation of alternative models capturing the essentials of the quantum plasma dynamics.

In this context, recently \cite{Haas} a fluid moments hierarchy was derived from the electrostatic Wigner equation. As usual in moments theories \cite{Grad}, a set of macroscopic variables (particle density, current etc.) were defined in terms of integrals of the Wigner function. The time-evolution of these quantities was then deduced from the Wigner equation. No assumptions were made on the particular local equilibrium Wigner function. In the linear limit, a quantum version of the Bohm-Gross dispersion relation was derived. Also certain nonlinear traveling wave solutions were obtained. 

It is the central purpose of this work, to extend the results of Ref.\ \cite{Haas} to the electromagnetic case. Hence we define the moments 
\begin{eqnarray}
\label{ee21}
&& \!\!\!\! \!\!\!\! n = \int\! d{\bf v} f \,,\\
\label{ee22}
&& \!\!\!\! \!\!\!\! n{\bf u} = \int\! d{\bf v}\,f\, {\bf v} \,,\\
\label{ee23} 
&& \!\!\!\! \!\!\!\! P_{ij} = m\,\left(\int\! d{\bf v}\,f\,v_i\, v_j - n\,u_i\,u_j\right) \,, \\
\label{ee24}
&& \!\!\!\! \!\!\!\!Q_{ijk} = m\,\int\! d{\bf v}\,(v_i - u_i) (v_j - u_j) (v_k - u_k)\,f \,, \\
\label{ee25}
&& \!\!\!\! \!\!\!\!R_{ijkl} = m\,\int\! d{\bf v}\,(v_i - u_i) (v_j - u_j) (v_k - u_k) (v_l - u_l)\,f \,,
\end{eqnarray}
and so on, as if $f$ were a classical distribution function. Since all quantities are postulated in a gauge invariant way we can safely interpret $n$, ${\bf u}$, $P_{ij}$ etc. respectively as a particle density, a velocity field, a second rank stress tensor and so on. In particular, a scalar pressure $p = (1/3)\,P_{ii}$ and a heat flux vector $q_i = (1/2)\,Q_{jji}$ can be deduced, where the summation convention is employed. Now the task is to obtain from the Wigner equation the equations of motion for the several moments, which will compose an infinite coupled hierarchy. 

We also note that for the case of an isotropic distribution function, \textit{i.e.}\ dependence of $f$ on the magnitude of the velocity only, all the odd moments must vanish from symmetry constraints, while the even moments are expressible in scalar quantities (by decomposition in terms of $\delta_{ij}$). Moreover, for the case of local rotational symmetry, \textit{i.e.}\ the existence of one preferred direction in (say, $\hat{\textbf{z}}$) due to an external magnetic field or an initial temperature anisotropy, we have the the form 
\begin{equation}
	P_{ij} = P^\perp h_{ij} + P^\|\hat{z}_i\hat{z}_j .
\end{equation}
and 
\begin{equation}
	Q_{ijk} = Q^\perp h_{(ik}\hat{z}_{k)} + Q^\|\hat{z}_i\hat{z}_j\hat{z}_k ,
\end{equation}
and similarly for higher order moments. Here we have introduced the projection tensor $h_{ij} = \delta_{ij} - \hat{z}_i\hat{z}_j$. These algebraic forms also solves the constraint equations (see below) that occur when assuming a stationary and homogeneous (but possibly anisotropic) equilibrium distribution.\footnote{We note that we can always decompose a moment of any order into its irreducible parts by picking an arbitrary direction and forming the projection operator orthogonal to that direction.}

For the sake of calculating the moments hierarchy equations, it is convenient to expand $\Delta\tilde{\bf v}$, $\tilde{\bf B}$ and $\tilde{\bf E}$ according to 
\begin{eqnarray}
\label{e12}
\Delta\tilde{v}_i &=& - \frac{q\hbar^2\varepsilon_{ijk}}{12m^3}\partial_m B_k\frac{\partial^2}{\partial v_j\,\partial v_m} + \frac{q\hbar^4\varepsilon_{ijk}}{540m^5}\partial^3_{mnl} B_k\frac{\partial^4\quad\quad\quad}{\partial v_j\partial v_m\partial v_n\partial v_l} + \dots 
\,, \\
\label{e13}
\tilde{E}_i &=& E_i - \frac{\hbar^2}{24m^2}\partial^2_{jk}E_i\frac{\partial^2}{\partial v_j\,\partial v_k} + \frac{\hbar^4}{1920m^4}\partial^4_{jkmn} E_i\frac{\partial^4\quad\quad\quad}{\partial v_j\partial v_k\partial v_m\partial v_n} + \dots \,,\\
\label{e14}
\tilde{B}_i &=& B_i - \frac{\hbar^2}{24m^2}\partial^2_{jk} B_i\frac{\partial^2}{\partial v_j\,\partial v_k} + \frac{\hbar^4}{1920m^4}\partial^4_{jkmn} B_i\frac{\partial^4\quad\quad\quad}{\partial v_j\partial v_k\partial v_m\partial v_n} + \dots \,,
\end{eqnarray}
disregarding higher order quantum corrections. The notation $\partial_i \equiv \partial/\partial r_i$ is used whenever there is no risk of confusion. 

Assuming decaying boundary conditions, as far as the moment hierarchy is closed at the third-rank stress tensor, only the leading quantum corrections [the terms $\propto\,\hbar^2$ in Eqs.\ (\ref{e12})--(\ref{e14})] are needed. This is due to the structure of the higher order corrections. Indeed, these terms always involve at least fourth-order velocity derivatives and, for instance, 
\begin{equation}
\label{e15} 
\int\,d{\bf v}\,v_i\,v_j\,v_k\,\frac{\partial^4\,f}{\partial v_a\partial v_b\partial v_c\partial_d} = 0 \,.
\end{equation}
Therefore, only the semiclassical Wigner equation is needed, which does not mean that the quantum effects are necessarily small. It just happens that higher order quantum corrections would appear only for higher order moment evolution equations.

Following Eq.\ (\ref{e11}), the semiclassical electromagnetic Wigner equation then reads
\begin{eqnarray}
&& \fl 
\Bigl[\frac{\partial}{\partial t} + {\bf v}\cdot\frac{\partial}{\partial{\bf r}} + \frac{q}{m}\,\left({\bf E} + {\bf v}\times{\bf B} \right)\,\cdot\frac{\partial}{\partial{\bf v}}
\Bigr]\,f({\bf r},{\bf v},t)  \nonumber \\ 
&& \fl \qquad = \frac{q\hbar^2}{24m^3}\partial^2_{jk}E_i\frac{\partial^3\,f}{\partial v_i\partial v_j\partial v_k} + \frac{q\hbar^2\varepsilon_{ijk}}{12m^3}\partial_m B_k\frac{\partial^3 f}{\partial r_i\partial v_j\,\partial v_m} + \frac{q\hbar^2\varepsilon_{ijk}v_j}{24m^3}\partial^2_{mn} B_k\frac{\partial^3 f}{\partial v_i\partial v_m\,\partial v_n} \label{e16} \\
&& \fl \qquad\qquad + \frac{q^2\hbar^2}{12m^4}\left(B_i\partial_j B_k\frac{\partial^3\,f}{\partial v_i\partial v_j\partial v_k} - 
B_i\partial_j B_i\frac{\partial^3\,f}{\partial v_j\partial v_k\partial v_k}\right) \,. \nonumber 
\end{eqnarray}
Notice that apparently the semiclassical electromagnetic Wigner equation, which has some interest in itself, was not discussed before in the literature. 

Calculating the moments, the result is 
\begin{eqnarray}
\label{e17}
&& \fl 
\frac{D\,n}{D\,t} + n\nabla\cdot{\bf u} \quad = \quad 0 \,, \\
\label{e18}
&& \fl 
\frac{D\,u_i}{D\,t} = - \frac{\partial_j P_{ij}}{m n} + \frac{q}{m}\,\left({\bf E} + {\bf u}\times{\bf B} \right)_i \,, \\
&& \fl 
\frac{D P_{ij}}{D\,t} = - P_{ik}\,\partial_{\,k} u_{j} - P_{jk}\,\partial_{\,k} u_{i}  - P_{ij}\nabla\cdot{\bf u} + \frac{q}{m}\varepsilon_{imn}P_{jm}B_n + \frac{q}{m}\varepsilon_{jmn}P_{im}B_n   \nonumber \\
\label{e19}
&& \fl \qquad + \frac{q\hbar^2}{12m^2}\varepsilon_{ikl}\partial_{l}\left(n\partial_{j}B_{k}\right)+ \frac{q\hbar^2}{12m^2}\varepsilon_{jkl}\partial_{l}\left(n\partial_{i}B_{k}\right)
- \partial_{\,k} Q_{ij\,k} \,, \\
&& \fl 
\frac{D\,Q_{ijk}}{D\,t} = - Q_{ijr}\,\partial_{\,r} u_{k} - Q_{jkr}\,\partial_{\,r} u_{i} - Q_{kir}\,\partial_{\,r} u_{j}  
- Q_{ijk}\nabla\cdot{\bf u} - \partial_{\,r} R_{ij\,kr} \nonumber \\ 
&& \fl \qquad + \frac{1}{mn}\Bigl(P_{ij}\partial_{r}P_{kr} + P_{jk}\partial_{r}P_{ir} + P_{ki}\partial_{r}P_{jr}\Bigr) + \frac{q}{m}\Bigl(\varepsilon_{irs}Q_{rjk} + \varepsilon_{jrs}Q_{rki} + \varepsilon_{krs}Q_{rij}\Bigr)B_s \nonumber \\
&& \fl \qquad - \frac{q\hbar^2 n}{12m^2}\Bigl(\partial_{ij}^{2}E_k + \partial_{jk}^{2}E_i + \partial_{ki}^{2}E_j\Bigr) + \frac{q^2\hbar^2 n}{12m^3}\Bigl(\delta_{ij}\partial_k + \delta_{jk}\partial_i + \delta_{ki}\partial_j\Bigr)\,B^2 \nonumber \\ 
&& \fl \qquad - \frac{q\hbar^2 n}{12m^2}\Bigl[ ({\bf u}\times\partial_{jk}^{2}{\bf B})_i + ({\bf u}\times\partial_{ki}^{2}{\bf B})_j + ({\bf u}\times\partial_{ij}^{2}{\bf B})_k\Bigr] \label{e20} \\ 
&& \fl \qquad + \frac{q\hbar^2\,n}{12m^2}\Bigl[ \varepsilon_{irs}\left(\partial_{j}B_r\partial_{s}u_k + \partial_{k}B_r\partial_{s}u_j\right) + \varepsilon_{jrs}\left(\partial_{k}B_r\partial_{s}u_i + \partial_{i}B_r\partial_{s}u_k\right) \nonumber \\ 
&& \fl \qquad\qquad + \varepsilon_{krs}\left(\partial_{i}B_r\partial_{s}u_j + \partial_{j}B_r\partial_{s}u_i\right)\Bigr] 
  - \frac{q^2\hbar^2 n}{12m^3}\Bigl[\partial_{i}(B_j B_k) + \partial_{j}(B_k B_i) + \partial_{k}(B_i B_j)\Bigr] \,. \nonumber 
\end{eqnarray}
When ${\bf B} = 0$, Eqs.\ (\ref{e17}--\ref{e20}) recover the electrostatic equations \cite{Haas}. In the limit $\hbar \rightarrow 0$ it reproduce the classical electromagnetic moment hierarchy equations \cite{Goswami, Siregar,Ramos}. Quantum effects are explicit already in the transport equation for the pressure dyad, through the magnetic field. 

Previous approaches \cite{Gardner} derived quantum transport equations for charged particle systems assuming a local semiclassical Wigner function corresponding to a perturbed Maxwell-Boltzmann equilibrium. Here, however, the treatment include magnetic fields and is not semiclassical. A further approach for the derivation of quantum effects in macroscopic equations is through the eikonal decomposition of the wave functions of the quantum statistical ensemble and adequate simplifying assumptions \cite{Manfredi}. In both cases \cite{Gardner, Manfredi} the pressure dyad $P_{ij}$ would be expressed as the sum of a classical part and a quantum part, the later one associated to a Bohm potential term in the force equation (\ref{e18}).

If we have used the gauge dependent Wigner function, it would not be possible to proceed exactly as in the classical case in the definition of the moments. Indeed, it would be natural to postulate them as 
\begin{eqnarray}
\label{x1}
&&  n = \int\! d{\bf p} f^{\,GD} \,,\\ \label{x2}
&&  n{\bf u} = \int\! d{\bf p}\, \left(\frac{{\bf p} - q\,{\bf A}}{m}\right) \,f^{\,GD} \,,\\
\label{e21}
&& P_{ij} = m\,\left(\int\! d{\bf p}\,\frac{(p_i - q\,A_i)(p_j - q\,A_j)}{m^2}\,f^{\,GD} - n\,u_i\,u_j\right) \,,\\
\label{e22}
&& Q_{ijk}^{GV} = \frac{1}{m^2}\int\! d{\bf p}(p_i - qA_i - mu_i) (p_j - qA_j - mu_j) (p_k - qA_k - mu_k)f^{GV} \,.
\end{eqnarray}
The same symbols $n$, ${\bf u}$ and $P_{ij}$ are used on purpose since Eqs.\ (\ref{x1})--(\ref{e21}) produce the same expressions as from the GIWF, in spite of the fact that $f^{GV}$ itself is a gauge dependent object. However, from the equation satisfied by the usual Wigner equation \cite{Haas2} one would obtain  
\begin{eqnarray}
\nonumber 
&& \frac{D P_{ij}}{D\,t} = - P_{ik}\,\partial_{\,k} u_{j} - P_{jk}\,\partial_{\,k} u_{i}  - P_{ij}\nabla\cdot{\bf u} + \frac{q}{m}\varepsilon_{imn}P_{jm}B_n + \frac{q}{m}\varepsilon_{jmn}P_{im}B_n \\ \label{e23} 
&& \qquad\quad  - \frac{q\,\hbar^2}{4\,m^2}\,\partial^{2}_{ij}{\bf A}\cdot\nabla n - \partial_{\,k} Q_{ij\,k}^{\,GV} \,,
\end{eqnarray}
containing gauge dependent quantum terms. The reason is that 
\begin{equation}
\label{y}
Q_{ijk}^{\,GV} = Q_{ijk} - \frac{q\,\hbar^2\,n}{12\,m^2}\left(\partial^{2}_{ij}\,A_k + \partial^{2}_{jk}\,A_i + \partial^{2}_{ki}\,A_j\right) 
\end{equation}
is not gauge invariant. If $Q_{ijk}^{\,GV}$ from Eq.\ (\ref{y}) is inserted into Eq. (\ref{e23}) one re-derive Eq.\ (\ref{e19}) for the pressure dyad on taking into account the Coulomb gauge which is assumed \cite{Haas2} in the evolution equation for $f^{\,GD}$. 

Similarly the transport equations for the higher order moments are not gauge invariant.  The conclusion is that to derive consistent equations from the usual Wigner function we would be obliged to modify the definition of moments. However, in this case there is the lost of one of the key advantages of using Wigner functions, namely the strict resemblance with the classical formalism. Also notice that if the heat flux tryad is set to zero the quantum term in Eq.\ (\ref{e23}) is nonlinear for unmagnetized homogeneous equilibria, unlike Eq.\ (\ref{e19}) where a quantum contribution survives in this situation. 

In principle one could use the gauge dependent Wigner function to consistently calculate the higher order moments such as $Q_{ijk}, R_{ijkl}$ and so on. However, due to the fact that operators in quantum mechanics in general are non-commuting this cannot be done in practice. To see how this comes about we consider calculating the second order moment using the gauge dependent Wigner function. Calculating the second order moment $P_{ij}(\mathbf r,t)$ involves finding the expectation value of the operator, given by%
\footnote{The definition of the pressure operator \,in quantum mechanics is motivated by considering the Heisenberg evolution equation for the probability current operator which will be coupled to the divergence of the pressure operator.} 
\begin{equation}
	\hat \Pi_{ij} = \frac{1}{4\,m} \left[ \hat p_i - q A_i (\hat{\mathbf r}, t) , 
	\left[ \hat p_j - q A_j (\hat{\mathbf r}, t) , \delta(\hat{\mathbf r} - \mathbf r) \right]_+  
	\right] _+ ,
\end{equation}
where $\left[ \hat a , \hat b \right]_+ = \hat a \hat b + \hat b \hat a$ denotes the anti-commutator. In order to calculate the expectation value using the Wigner formalism it is necessary to map the operator into a phase-space function using Weyl-correspondence \cite{Weyl}. This is done in practice by ordering the operators into a symmetric product of the position and momenta operators by using the commutation relations and then make the substitutions $\hat{\mathbf r} \rightarrow \mathbf r$ and $\hat{\mathbf p} \rightarrow \mathbf p$. It turns out that the correct phase-space function is obtained by just making the substitution in the operator above without first Weyl order it. Hence we may calculate the pressure dyad using the gauge dependent Wigner function as
\begin{equation}
\fl	P_{ij} (\mathbf r ,t) = \int \frac{d\mathbf r' d \mathbf p}{m} \left[ p_i - qA_i(\mathbf r',t) \right] 
	\left[ p_j - q A_j (\mathbf r',t) \right] \delta({\bf r}' - {\bf r}) f^{\,GD}({\bf r}',{\bf p},t)
	- m n u_i  u_j 
\end{equation}
However, for the third order moment $Q_{ijk}$ the correct phase-space function is not obtained simply by making the substitution $\hat{\mathbf r} \rightarrow \mathbf r$ and $\hat{\mathbf p} 
\rightarrow \mathbf p$. Hence, calculating the correct third order moment using the gauge dependent Wigner function is complicated and involves Weyl ordering the corresponding operator so as to obtain the correct phase-space function. 

The GIWF has a modified Weyl ordering rule, discussed in \cite{Serimaa} and calculating the moments is done in complete analogy with the classical case, see Eqs.\ (\ref{ee21})--(\ref{ee25}). 

\section{Transverse dispersion relation}

As an application of the fluid Eqs.\ (\ref{e17}--\ref{e20}) we now consider linear transverse waves. Considering an one-component plasma, where the ions acts only as an homogeneous neutralizing background with number density $n_0$, the moment equations can be linearized around the equilibrium  
$n = n_0, {\bf u} = 0, P_{ij} = P_{ij}^{(0)}, Q_{ijk} = 0, R_{ijkl} = 0, {\bf E} = 0, {\bf B} = 0$. To consider waves propagating in the z-direction with transverse polarization we let all fluctuations have the space-time dependence $e^{ikz - i \omega t}$ and set $E_z = 0$. Moreover we decompose the zeroth order pressure dyad as $P_{ij}^{(0)} = P_{\perp} (\delta_{i x } \delta_{j x} + \delta_{i y} \delta_{j y} ) + P_{||} \delta_{i z} \delta _{jz}$, where $P_\perp$ and $P_\parallel$ are constants.

It turns out that if we use the closure assumption $R_{ijkl} = 0$ the quantum corrections to the transverse modes will not be retained so that to display the lowest order quantum corrections it is necessary to take into account also the contribution from the fourth order moment. As a closure assumption we use 
\begin{equation}
\label{c}
R_{ijkl} = - \frac{q\hbar^2}{4m^3\omega^2} \left( P^{(0)}_{im} \partial_{jkl}^3 
	+ P^{(0)}_{jm} \partial_{kli}^3 + P^{(0)}_{km} \partial_{lij}^3 + P^{(0)}_{lm} \partial_{ijk}^3 \right)\,E_m \,,
\end{equation}
adapted to the transverse wave case. The closure (\ref{c}) is deduced systematically from the linearized equations satisfied by the fourth and fifth order moments, see Appendix A. Note that in principle the fourth order moment $R_{ijkl}$ can have a nonzero equilibrium contribution $R^{(0)}_{ijkl} \sim v_T^4$, where $v_T = \sqrt{(2\,P_\perp + P_{||})/(m\,n_0)}$ is the thermal velocity, but we will neglect this since we are looking only for the lowest order correction. Likewise for the terms $\sim \hbar^4$. Finally, it is worth to remark that in the classical limit the fourth order moment could be set to zero. 

The linearized equations can then be solved by first writing the magnetic field in terms of the electric field and then eliminating all quantities except the velocity so that we obtain the velocity in terms of the electric field. 
Coupling the resulting equation with Faraday's law via the current density $\mathbf J = q n_0 \mathbf u$ the dispersion relation
\begin{eqnarray}
	\omega^2 - k^2 c^2 
	&=&
	\omega_p^2\left[ 1 + \frac{k^2 P_\perp}{n_0 m \omega^2} 
	+ \frac{\hbar^2 k^6 P_\perp}{4 n_0 m^3 \omega^4} \right] ,
	\label{fluiddisp}
\end{eqnarray}
is obtained. Here $\omega_p = \sqrt{n_0\, q^{2}/(m \,\epsilon_0)}$ is the plasma frequency. If, instead, the closure $R_{ijkl}= 0$ was used, the term proportional to $\hbar^2$ would be absent in the dispersion relation. 

In the simultaneous long wavelength and semiclassical limits, Eq. (\ref{fluiddisp}) can be shown to admit an approximate solution
\begin{equation}
\label{d}
\omega^2 \simeq \omega_{p}^2 + c^2\,k^2 + \frac{P_\perp\,k^2}{m\,n_0} + \frac{\hbar^2\,k^6\,P_\perp}{4\,m^3\,n_0\,\omega_{p}^2} \,.
\end{equation}

To check the consistency, we need to compare to the results from kinetic theory. Here we are not concerned with Landau damping issues so that all integrals can be interpreted in the principal value sense. Assume 
\begin{eqnarray}
\label{e24}
&& {\bf E} = {\bf E}_{1}\,\exp[i(k\,z-\omega\,t)] \,,\\
\label{e25}
&& {\bf B} = {\bf B}_{1}\,\exp[i(k\,z-\omega\,t)] \,,\\
\label{e26}
&& f = f_{0}({\bf v}) + f_{1}({\bf v})\,\exp[i(k\,z-\omega\,t)] \,,
\end{eqnarray}
where ${\bf k}\cdot{\bf E} = 0$ as before and 
with the subscript 1 denoting first order quantities. The equilibrium Wigner function satisfy
\begin{equation}
\label{e27}
\int\,d{\bf v}\,f_0 = n_0 \,, \quad \int\,d{\bf v}\,{\bf v}\,f_0 = 0 \,.
\end{equation}
Further we assume an equilibrium Wigner function such that $f_0 = f_{0}(v_\perp, v_z)$, where $v_{\perp}^2 = v_{x}^2 + v_{y}^2$.
Notice that since there is no zeroth order magnetic field the perturbation velocity $\Delta\tilde{\bf v}$ is also of first order. Hence $\Delta\tilde{\bf v}$ does not contribute in the linearized Wigner equation (\ref{e11}). Using Eqs.\ (\ref{e9}--\ref{e10}) we get 
\begin{equation}
\label{e28}
\tilde{\bf E} = {\bf E}\,L \,,\quad \tilde{\bf B} = {\bf B}\,L \,,
\end{equation}
defining the operator 
\begin{equation}
\label{e29}
L = \frac{\sinh\theta}{\theta} \,, \quad \theta = \frac{\hbar\,k}{2\,m}\frac{\partial}{\partial\,v_z} \,.
\end{equation}
We note that 
\begin{equation}
\label{e30}
L\left(\frac{\partial f_0}{\partial\,v_z}\right) = \frac{m}{\hbar\,k}\,\left[f_{0}\left({\bf v} + \frac{\hbar\,{\bf k}}{2\,m}\right) - f_{0}\left({\bf v} - \frac{\hbar\,{\bf k}}{2\,m}\right)\right] \,,
\end{equation}
where ${\bf k} = k\,{\bf\hat{z}}$. Moreover $L \rightarrow 1$ in the classical limit, since
\begin{equation}
\label{e31}
L = \sum_{j=0}^{\infty}\,\frac{1}{(2\,j+1)!}\left(\frac{\hbar\,k}{2\,m}\,\frac{\partial}{\partial\,v_z}\right)^{2\,j} = 1 + \frac{1}{24}\left(\frac{\hbar\,k}{m}\right)^2\,\frac{\partial^2}{\partial\,v_{z}^2} + \dots
\end{equation}

Then linearizing the Wigner equation (\ref{e11}) and from the Maxwell equations with charge and current densities $q\,\left(\int\,d{\bf v}\,f - n_{0}\right)$ and $q\,\int\,d{\bf v}\,{\bf v}\,f$ respectively, the result is 
\begin{equation}
\label{e32}
\omega^2 = \omega_{p}^2 + c^2\,k^2 + \frac{k^2\,\omega_{p}^2}{2\,n_0}\,\int\,d{\bf v}\,\frac{v_{\perp}^2\,L\,f_0}{(\omega - {\bf k}\cdot{\bf v})^2} \,,
\end{equation}
where $c$ is the speed of light and $\omega_p$ is the plasma frequency. In comparison to the classical transverse dispersion relation, the only change is the replacement $f_0 \rightarrow \tilde{f}_0 = L\,f_0$. In a classical picture it is as if the particle velocities were reorganized through the diffusive operator $L$. Also notice that still $\tilde{f}_0 = \tilde{f}_{0}(v_\perp, v_z)$. Moreover, the quantum diffusion induced by the operator $L$ preserves the number of particles, since $\int\,d{\bf v}\,\tilde{f}_{0} = \int\,d{\bf v}\,{f}_{0}$ due to Eq. (\ref{e31}) under decaying boundary conditions. Figure 1 shows the effect of $L$ on the equilibrium $f_0 = f_{T}(v_{\perp})\,\exp[-v_{z}^2/(2\,v_{0}^2)]$, for different values of the non dimensional parameter $H = \hbar\,k/(2\,m\,v_0)$. 
\begin{figure}
\begin{center}
\includegraphics[width=0.8\textwidth]{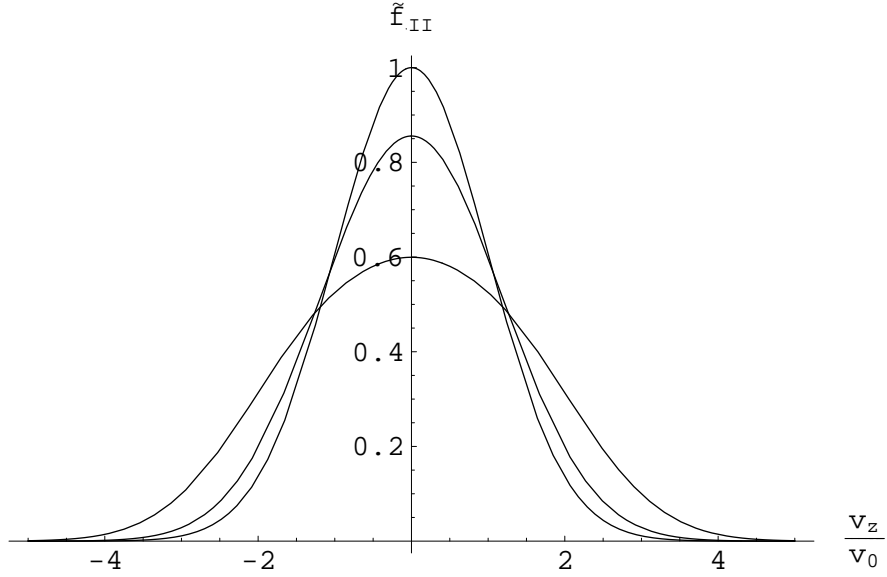}
\caption{Quantum diffusion on the equilibrium Wigner function $f_0 = f_{T}(v_{\perp})\,\exp[-v_{z}^2/(2\,v_{0}^2)]$. Here $\tilde{f}_{||} = L\,(\exp[-v_{z}^2/(2\,v_{0}^2)])$. Values of the parameter $H = \hbar\,k/(2\,m\,v_0)$ are $H = 0, 1$ and $2$, so that $\tilde{f}_{||}(0) = 1,\,\, 0.86$ and $0.60$ respectively.}
\label{figure1}
\end{center}
\end{figure} 
In the simultaneous long wavelength and semiclassical limits and retaining only the leading $\sim v_T^2$ thermal corrections, Eqs. (\ref{d}) and (\ref{e32}) give the same result via the natural identification $P_\perp = (m/2)\int d{\bf v}v_\perp^2 f_0$. This conclude the equivalence between the moments and kinetic theories, in the fluid limit. 

To compare,  the transverse dispersion relation following from the gauge dependent Wigner function \cite{Klimontovich, Kuzelev} can be expressed as 
\begin{equation}
\label{e33}
\omega^2 = \omega_{p}^2 + c^2\,k^2 - \frac{m\,\omega_{p}^2}{2\,n_0\,\hbar}\,\int\,d{\bf v}\,\frac{v_{\perp}^2}{\omega - {\bf k}\cdot{\bf v}}\left[ f_{0}\left({\bf v} + \frac{\hbar\,{\bf k}}{2\,m}\right) - f_{0}\left({\bf v} - \frac{\hbar\,{\bf k}}{2\,m}\right)\right] \,,
\end{equation}
or, using Eq. (\ref{e30}), as
\begin{equation}
\label{e34}
\omega^2 = \omega_{p}^2 + c^2\,k^2 - \frac{\omega_{p}^2\,k}{2\,n_0}\,\int\,d{\bf v}\,\frac{v_{\perp}^2}{\omega - {\bf k}\cdot{\bf v}}\,L\left(\frac{\partial f_0}{\partial\,v_z}\right) \,.
\end{equation}
An integration by parts then shows the equivalence with the gauge invariant transverse dispersion relation Eq.\ (\ref{e32}). Therefore the gauge choice issues tend to be crucial only for the nonlinear regimes, as also manifest in the gauge dependent nonlinear term in Eq.\ (\ref{e23}) for the pressure dyad. However, in the case of non-homogeneous equilibria the use of a gauge independent electromagnetic Wigner equation is advisable even for linear waves. 

\section{Conclusion}

The moment hierarchy equations derived from the GIWF electromagnetic evolution equation is obtained. The advantages over the gauge dependent Wigner formalism are stressed. Discrepancies tend to be prominent in the nonlinear regimes and for higher order moments of the Wigner function. The fluid-like equations (\ref{e17})--(\ref{e20}), closed at the transport equation for the heat flux tryad, is applied to the propagation of linear transverse waves. Good agreement is found when comparing with the results from kinetic theory, in the long wavelength approximation. A key ingredient to a successful macroscopic theory is an adequate closure of the moment equations and a recipe for solving this question is proposed, see Appendix A. The approach is not restricted to particular local equilibrium GIWFs and is not based on a Madelung decomposition of the quantum statistical ensemble wave functions. The moment equations (\ref{e17})--(\ref{e20}) is an adequate starting point for studying the nonlinear aspects of quantum plasma problems involving magnetic fields, {\it e.g.} via numerical simulations. 

\ack
F.H. acknowledges the support provided by Ume{\aa} University and the Kempe Foundations. 
This work is supported by the European Research Council under Contract No.\ 204059-QPQV, and the Swedish Research Council under Contract No.\ 2007-4422.

\appendix 
\section{The closure problem}
The closure (\ref{c}) can be deduced systematically from linearized higher order moment equations. Let $S_{ijklm}$ be the fifth order moment defined in analogy to the third and fourth order moments, see Eqs. (\ref{ee24}--\ref{ee25}). The sixth order moment will be set to zero. The evolution equations for the fourth and fifth order moments are derived following the same steps as when Eqs.\ (\ref{e17}--\ref{e20}) were derived starting from Eq.\ (\ref{e16}). Since they are quite complicated we here only include the linear terms, which gives
\begin{eqnarray}
&& \fl \partial_t R_{ijkl} = 
	- \frac{q\hbar^2}{12 m^3} \Bigl[\epsilon_{inm} \Bigl(P^{(0)}_{jn} \partial_{kl}^2 
	+ P^{(0)}_{kn} \partial_{lj}^2 + P^{(0)}_{ln} \partial_{jk}^2 + P^{(0)}_{jk} \partial_{ln}^2 + P^{(0)}_{kl} \partial_{jn}^2 + P^{(0)}_{lj} \partial_{kn}^2\Bigr) 
		\nonumber \\ 
	&& \fl \qquad + \epsilon_{jnm} \Bigl(P^{(0)}_{kn} \partial_{li}^2 +  P^{(0)}_{ln} \partial_{ik}^2
	+ P^{(0)}_{in} \partial_{kl}^2 + P^{(0)}_{kl} \partial_{in}^2 +  P^{(0)}_{li} \partial_{kn}^2 + P^{(0)}_{ik} \partial_{ln}^2 \Bigr) \nonumber \\ 
	&& \fl \qquad + \epsilon_{knm} \Bigl( P^{(0)}_{ln} \partial_{ij}^2 + P^{(0)}_{in} \partial_{jl}^2  
	+ P^{(0)}_{jn} \partial_{li}^2 + P^{(0)}_{ij} \partial_{ln}^2 +  
	P^{(0)}_{jl} \partial_{in}^2 + P^{(0)}_{li} \partial_{jn}^2 \Bigr) \nonumber 
	 \\  
	 && \fl \qquad + \epsilon_{lnm} \Bigl( P^{(0)}_{in} \partial_{jk}^2 +  
	P^{(0)}_{jn} \partial_{ki}^2 + P^{(0)}_{kn} \partial_{ij}^2 + P^{(0)}_{ij} \partial_{kn}^2 + P^{(0)}_{jk} \partial_{in}^2 + P^{(0)}_{ki} \partial_{jn}^2\Bigr) \Bigr] B_m \nonumber \\
	  \label{r} 
	  && \fl \qquad - \partial_m S_{ijklm} \,, \\
	  && \fl
	  \partial_t S_{ijklm} =  - \frac{q \hbar^2}{12 m^3} 
		\Bigl[
			\Bigl( 
				P^{(0)}_{ij} \partial_{kl} + P^{(0)}_{jk} \partial_{li}  + P^{(0)}_{kl} \partial_{ij} + P^{(0)}_{li} \partial_{jk}
				 + P^{(0)}_{ik} \partial_{jl} + P^{(0)}_{jl} \partial_{ik} 
			\Bigr)E_m
				\nonumber \\ 
	&& \fl \qquad + \Bigl( 
				P^{(0)}_{jk} \partial_{lm} +  P^{(0)}_{kl} \partial_{mj}  + P^{(0)}_{lm} \partial_{jk} + P^{(0)}_{mj} \partial_{kl}
				 + P^{(0)}_{jl} \partial_{km} + P^{(0)}_{km} \partial_{jl}			
			\Bigr)E_i \nonumber \\			
	&& \fl \qquad + \Bigl(P^{(0)}_{kl} \partial_{mi} +  P^{(0)}_{lm} \partial_{ik} + P^{(0)}_{mi} \partial_{kl} +
				P^{(0)}_{ik} \partial_{lm} 
				  + P^{(0)}_{km} \partial_{li} + P^{(0)}_{li} \partial_{km}
			\Bigr) E_j \nonumber  \\		 
	&& \fl \qquad + \Bigl(P^{(0)}_{lm} \partial_{ij} + P^{(0)}_{mi} \partial_{jl} +
				P^{(0)}_{ij} \partial_{lm} + 
				+ P^{(0)}_{jl} \partial_{mi} + P^{(0)}_{li} \partial_{mj} + P^{(0)}_{mj} \partial_{li}
			\Bigr) E_k
		\nonumber  \\		
	&& \fl \qquad + \Bigl(P^{(0)}_{mi} \partial_{jk} + 
				P^{(0)}_{ij} \partial_{km} + P^{(0)}_{jk} \partial_{mi} + P^{(0)}_{km} \partial_{ij}
				+ P^{(0)}_{mj} \partial_{ik} +  P^{(0)}_{ik} \partial_{mj}
			\Bigr) E_l\Bigr] \label{s}
		  \,. 
		 \end{eqnarray}
After Fourier transforming and inserting $S_{ijklm}$ from Eq. (\ref{s}) into Eq. (\ref{r}), Eq. (\ref{c}) is derived using Faraday's law. The procedure is adapted to the present equilibrium (homogeneous, no streaming particles, no heat flux, negligible higher order thermal effects). 

It turns out that due to a cancellation arising from Faraday's law in the transverse case, the result in Eq. (\ref{c}) is correct even if Eqs. (\ref{r}--\ref{s}) were extended to include $\sim \hbar^4$ terms. To obtain the next order quantum effects dispersion relation using the fluid theory, it is hence necessary to include higher order moments. In this example, the sixth order moment disregarded in Eq. (\ref{s}).

From the above we can infer a general recipe for the closure of the fluid-like system up to the $N^{th}-$moment: Fourier transform the linearized evolution equations for the $(N+1)^{th}$ and $(N+2)^{th}$ moments, setting the $(N+3)^{th}$ moment to zero. In this way we derive an expression for the $(N+1)^{th}$ moment, so as to close the system for the $N$ moments. The form of the linearized equations depends on the particular equilibrium. Na\"{i}ve closures like setting the $(N+1)^{th}-$moment directly to zero tend to produce fake results when comparing to kinetic theory. This is in sharp contrast to the simplicity of the electrostatic case, where faithful equations are obtained already defining the fourth order moment to be zero \cite{Haas}.

\end{document}